# Microstructure, hardness and mechanical properties of two different unalloyed tantalum wires deposited via Wire + Arc Additive Manufacture

G. Marinelli*[,a], F. Martina[a], S. Ganguly[a], S. Williams[a]

[a]Welding Engineering and Laser Processing Centre (WELPC), College Road, Cranfield University, Cranfield, MK43 0AL, UK

*Corresponding Author. E-mail address: g.marinelli@cranfield.ac.uk (G. Marinelli)

**Abstract**

An innovative way of producing large-scale unalloyed tantalum parts, based on the Wire + Arc Additive Manufacturing process, has been developed in this study. Two different unalloyed tantalum wires have been used to deposit 200-mm-long structures in tantalum. The effect of the wire chemistry on microstructure, hardness, porosity, mechanical properties and strain localisation has been investigated. The deposits showed high integrity and excellent mechanical properties, with yield strength, ultimate tensile strength and elongation as high as 234 MPa, 261 MPa, and 36 %, respectively. Indeed, yield strength was higher than commercially available tantalum, even though, in this study, the grains were large and had a high aspect ratio. Wire + Arc Additive Manufacture has clearly shown the potential to produce tantalum components with relatively low cost and reduced lead time, thus offering a new robust and viable manufacturing route.

## 1   Introduction

Additive Manufacturing (AM) has been identified as a very promising technology with the capability to reinvigorate and remodel the manufacturing sector [1,2]. The technology is based on the simple concept of depositing three-dimensional structures using a layer-by-layer approach [3]. Cost and time reduction, design freedom and engineered materials properties are the main business benefits which AM can enable [3].

Among the different AM techniques, wire-feed technologies and in particular the Wire + Arc Additive Manufacturing (WAAM) process, which employs an electric arc as the heat source, have already proven capable of producing large-scale components [4,5]. WAAM can directly fabricate fully-dense metallic large 3-D near-net-shape components with a much higher deposition rate, than most other metal additive manufacture processes [4,5], the highest rate so far being of 9.5 kg/h [6]. The WAAM process has successfully produced large-scale parts in stainless steel [7], Inconel ® [8], titanium [9], aluminium [10] and tungsten [11]. Furthermore, functionally graded structures of refractory metals have also been deposited using WAAM [12]. The manufacture of large and engineered components by WAAM is attractive also because of the low system and operating costs, as well as the modularity of the system design [5,13].

With a melting point of 3017°C and a density of 16.6 g cm$^{-3}$, tantalum is a useful material for specialised applications because of its stability at high temperature, high corrosion resistance and high melting point [14]. Unlike other refractory metals, such as tungsten and molybdenum, tantalum is characterised by a high ductility at room temperature [15,16]. For these reasons, tantalum finds its utilisation within a large range of applications, when a combination of all these particular properties is required. Indeed, tantalum's main areas of use are within the defence sector as well as the electronics, aerospace and chemical ones[14,15,17].

The development of AM technologies suitable for the deposition of unalloyed tantalum could lead to a larger demand for components within many industries with considerable cost reduction. Different AM studies, in which lasers were used as the heat source and tantalum powder as the feedstock, have already been reported. In particular, a complete study on the Laser-powder-bed-fusion (LPBF) process, microstructure and mechanical properties of tantalum powder has been conducted by Zhou et al. [18]. Micro-pores and micro-cracks were found within the build when the laser power was not optimised; both high hardness and high tensile properties were observed after the deposition. A large loss in ductility was also reported, accompanied by a brittle-like fracture behaviour. It is important to note that the raw materials used in this study had a concentration of 0.18 wt.% of oxygen. These oxygen levels are considerably high, and it is known that oxygen leads to strengthening in the tantalum components [19]. In the work of Thijs et al. [20], the LPBF process was studied for tantalum, focusing mainly on the microstructural evolution and mechanical properties. Epitaxial grain growth through the layers was observed and led to the formation of large columnar grains and the development of strong texture. The microstructure's anisotropy was also reflected in the mechanical properties. A further study on porous tantalum parts is reported in the study of Wauthle et al. [21], in which LPBF was used effectively to produce porous tantalum implants with fully-interconnected open pores. Finally, a tantalum coating was successfully deposited onto titanium using a different AM process, namely Laser-engineered net shaping (LENS), to produce controlled porous components with enhanced bio-properties [22].

Unlike other BCC metals, tantalum shows a high solubility for interstitials, which hardly segregate at the grain boundaries. This is one of the main reasons for its improved ductility at room temperature [14,16]. However, the mechanical properties of tantalum are highly influenced by the purity, the temperature and the strain rate [16,17,23]. In particular, interstitials such as nitrogen and oxygen have a strong effect on hardness and yield strength. The manufacturing route has also a large effect on mechanical properties, as shown in **Table 1**. Commercially unalloyed tantalum components produced using electron beam melting and powder metallurgy have similar mechanical properties; these are mainly influenced by the grain size. It is also clear that cold work impacts predominantly the hardness and the elongation (**Table 1**). The data reported for LPBF depict unusual properties that are predominantly influenced by the content of oxygen, up to date.

**Table 1**: Overview of mechanical properties of pure Ta obtained via different manufacturing routes. Data taken from the comparison reported in the work of Thijs et al. [20] and completed with the results from the work of Zhou et al. [18]. (*EB: electron beam melted; **P/M: powder metallurgy.)

| Ta Grade | Commercial pure EB* | Commercial pure P/M** | Soft annealed | Cold-worked | Laser-powder-bed-fusion |
|---|---|---|---|---|---|
| Modulus of Elasticity [GPa] | 185 | 185 | 186 | 186 | na |
| Yield Strength [Mpa] | 165 | 220 | na | na | 450 |
| Ultimate Tensile Strength [MPa] | 205 | 310 | 200-390 | 220-1400 | 739 |
| Elongation [%] | 40 | 30 | 20-50 | 2-20 | 2 |
| Vickers Hardness [HV] | 110 | 120 | 60-120 | 105-200 | 425 |

Most of the research reported sought the production of small-size complex tantalum structures for medical applications. To the authors' best knowledge, there are currently no studies either on the production of large-scale unalloyed tantalum components using AM or on the application of WAAM to tantalum. Thus, no mechanical properties are available for tantalum components produced via WAAM. In the present study, the impact of the chemistry of the two different wires on the evolution of porosity, microstructure, hardness and the mechanical properties of the deposited structures has been analysed as well.

## 2 Experimental Procedure

Two different unalloyed tantalum wires with a diameter of 1.2 mm were used for the WAAM process, for simplicity called Wire A and Wire B. Cold-rolled tantalum plates with dimensions of 210 mm in length, 50 mm in width and 4 mm in thickness were used for the initial trials of deposition, and for the production of the samples A1 and B1 used for the investigation of the microstructure and hardness (**Table 2**). Larger plates with dimensions of 400 mm in length, 60 mm in width and 8 mm in thickness were used for the deposition of the samples A2 and B2 and tensile coupons have been extracted from these (**Table 2**). The surface of every plate was ground and rinsed with acetone prior to the deposition, in order to remove most of the contaminants. **Fig. 1** shows the layout of the apparatus used in this study.

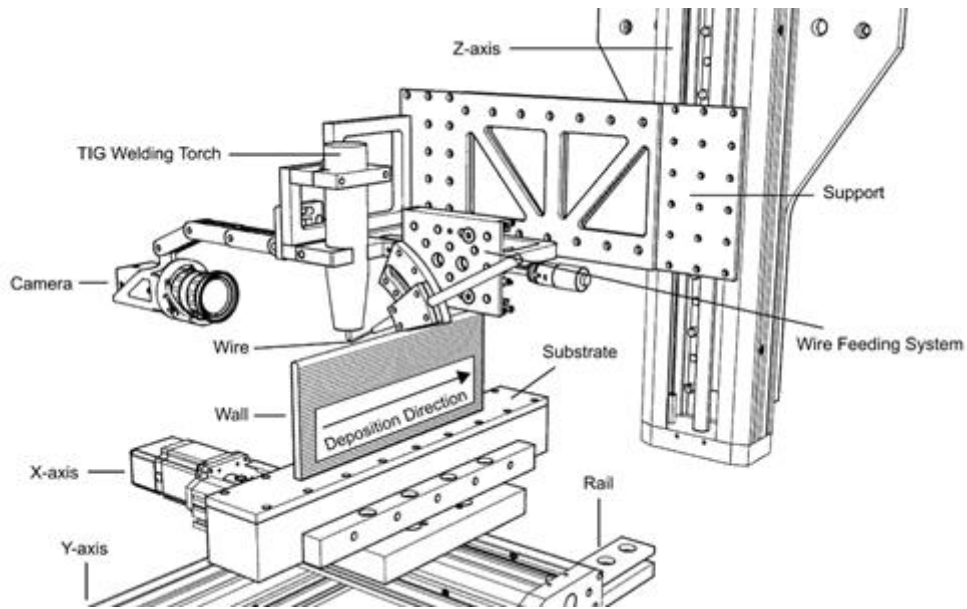

**Fig. 1**: Set-up used for the development of the WAAM process for unalloyed tantalum.

A conventional tungsten inert gas (TIG) welding torch, a power supply and a controlled wire feeder were used for the deposition. The heat source, the wire delivery system and the substrate were attached to three linear motorized high-load stages assembled in XYZ configuration. The tantalum wires were deposited layer-by-layer onto the substrate using a single bead. The direction of deposition was always kept constant for each successive layer, and the wire was always fed from the side of the weld pool. The apparatus was surrounded by an enclosure used to ensure a level of around 100 ppm of $O_2$ when purged with argon. A welding camera was used to study and monitor the process during the deposition of each layer. The camera was located opposite to the direction of wire feeding as shown schematically in **Fig. 1**.

**Table 2** reports a summary of the samples produced and analysed within this study. In particular, four straight walls were produced and analysed. They were labelled as A1, B1, A2 and B2. The first two were walls made of 20 layers each. The last two walls were produced with the dimensions of 230 mm in length, 110 mm in height and 10 mm in thickness by depositing 90 layers. Each structure was deposited using the same wire within the entire build (**Table 2**). The walls were produced using the parameters shown in **Table 3**. The parameters were kept constant throughout the entire process.

The microstructure was examined using a cross-section perpendicular to the deposition direction (extracted from the region reported in purple in **Fig. 2a**). The samples were polished and etched prior to the microstructural analysis. The surface discolouration was investigated using a scanning electron microscope (SEM) operating at 20 kV electron beam power. Vickers microhardness was measured using an automatic hardness testing machine. The main parameters for the hardness acquisition were 500g load and 10 seconds indentation time, for each testing point. A three-line scan with 0.5 mm spacing and thirteen points each was performed through height and the resulting values have been averaged for A2, B2 and Substrate. Furthermore, an analysis of the hardness at the interface between substrate and wall was conducted only for A1 and B1.

**Table 2:** Summary of the samples produced indicating the number of layers, the parental wire and the analysis performed for each sample.

| Sample name | Number of layers | Wire used | Analysis |
|---|---|---|---|
| A1 | 20 | Wire A | Hardness interface substrate/wall, geometrical control, microstructure. |
| B1 | 20 | Wire B | Hardness interface substrate/wall, geometrical control, microstructure. |
| A2 | 90 | Wire A | Chemical analysis, bead appearance, porosity, microstructure, tensile properties |
| B2 | 90 | Wire B | Chemical analysis, bead appearance, porosity, microstructure, tensile properties |

**Table 3**: WAAM process parameters used for the deposition of unalloyed tantalum.

| Parameter | Value |
|---|---|
| Travel Speed (TS) [mm/s] | 4 |
| Welding Current (I) [A] | 300 |
| Wire Feed Speed (WFS) [mm/s] | 40 |
| Shielding Gas Composition (SGC) [%] | 100 He |
| Gas Flow Rate (GFR) [L/min] | 15 |
| Oxygen Level [ppm] | ≈100 |

Six tensile coupons were extracted parallel and perpendicular to the build direction, from both sample A2 and B2, as shown in **Fig. 2a**. In agreement with the requirements of BS EN ISO 6892-1-2009 standard, and as shown in **Fig. 2b**, the specimens had a dog-bone shape with a gauge length of 27 mm, a gauge width of 6 mm and a thickness of 4 mm, resulting in a cross-sectional area of 24 mm². The overall length of the specimen was 89 mm and the radius of the fillet was 12 mm. Length and width of the grip section were both equal to 20 mm. Four further tensile specimens were extracted from a commercially available tantalum plate (of nature identical to that used as substrate) to provide a baseline for comparison. Each tensile coupon's surface was sprayed using first a white paint and then black paint, in order to create a speckle pattern for the measurement of displacement using digital image correlation (DIC).

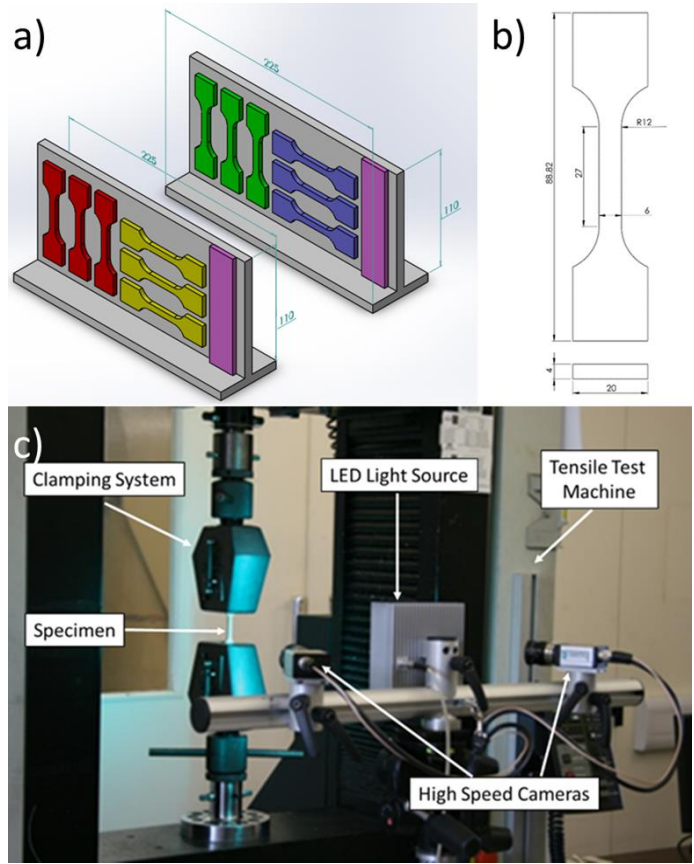

**Fig. 2**: (a) Extraction position and direction of the tensile coupons for each wall; (b) Main dimensions of the tensile coupon used for the tests; (c) Apparatus used for tensile tests and DIC recording.

Tensile tests were carried out with an Instron 5500R electromechanical machine equipped with a 20kN load cell. No extensometer has been used during testing as the displacement was calculated from the DIC data. The cross-head displacement speed was 1.0 mm/min. Two high-speed cameras were set up in the front of the sample on a stable tripod, coupled with a LED light source, as shown in **Fig. 2c**. LED illumination was used to overcome interference from the surrounding light. The cameras were equipped with a filter in order to capture and analyse only the reflected green light. The cameras were connected to a computer and the data was recorded using the software ISTRA 4D from Dantec Dynamics. The same software has been used for the post-processing of the digital image correlation and for the calculation of the yield strength, ultimate tensile strength and elongation.

Within the deposited samples A2 and B2, the parent welding wires and the substrates, oxygen and nitrogen levels, as well as other main metallic and non-metallic contaminants, were determined via chemical analysis.

## 3   Results and Discussion

*3.1   Chemical analysis*

**Table 4** shows the results of the chemical analysis of the substrate, the two wires and the two walls used for mechanical testing. All the materials used were high-purity unalloyed tantalum products. The substrate had the lowest content of nitrogen and oxygen respectively with a concentration of <10 ppm and 60 ppm. The two wires mainly differ in the oxygen level. This is reflected directly in the composition of the two walls built. Considered the content of oxygen in the deposition atmosphere was around 100 ppm, the content of oxygen only slightly increased, in both cases. More relevant is the difference of 62 ppm of oxygen between the two walls. The effect on the hardness and on the porosity is reported later.

**Table 4**: Elemental composition (wt.%) of the Substrate, Wire A, Wire B, A2 and B2.

|  | W | Mo | Ta | Ti | V | Cr | Fe | C | N | O | K |
|---|---|---|---|---|---|---|---|---|---|---|---|
| **Substrate** | <0.05 | <0.05 | 99.99 | <0.05 | <0.05 | <0.05 | <0.05 | 33 ppm | <10 ppm | 60 ppm | <10 ppm |
| **Wire A** | <0.05 | <0.05 | 99.98 | <0.05 | <0.05 | <0.05 | <0.05 | 36 ppm | 11 ppm | 190 ppm | <10 ppm |
| **Wire B** | <0.05 | <0.05 | 99.87 | <0.05 | <0.05 | <0.05 | <0.05 | 20 ppm | <10 ppm | 86 ppm | <10 ppm |
| **A2** | <0.05 | <0.05 | 99.97 | <0.05 | <0.05 | <0.05 | <0.05 | 48 ppm | 13 ppm | 226 ppm | <10 ppm |
| **B2** | <0.05 | <0.05 | 99.73 | <0.05 | <0.05 | <0.05 | <0.05 | 29 ppm | <10 ppm | 164 ppm | <10 ppm |

*3.2   Bead appearance*

**Fig. 3** shows the two walls deposited for the mechanical testing and their main dimensions, in particular, sample A2 and B2. Noteworthy is the complete absence of distortion after unclamping (**Fig. 3b**). This is mainly due to two factors: the high geometrical stiffness of the wall-plus-substrate component, and a localised plastic deformation of both substrate and the lower part of the wall which occurred while still clamped. From **Fig. 3a** and **Fig. 3b,** it can also be seen that both structures exhibit a more pronounced opacity on the lower surface, with respect to their top. The top layers were of a shiny silver appearance. These regions are shown with higher magnification in **Fig. 4a** and **Fig. 4b**, respectively.

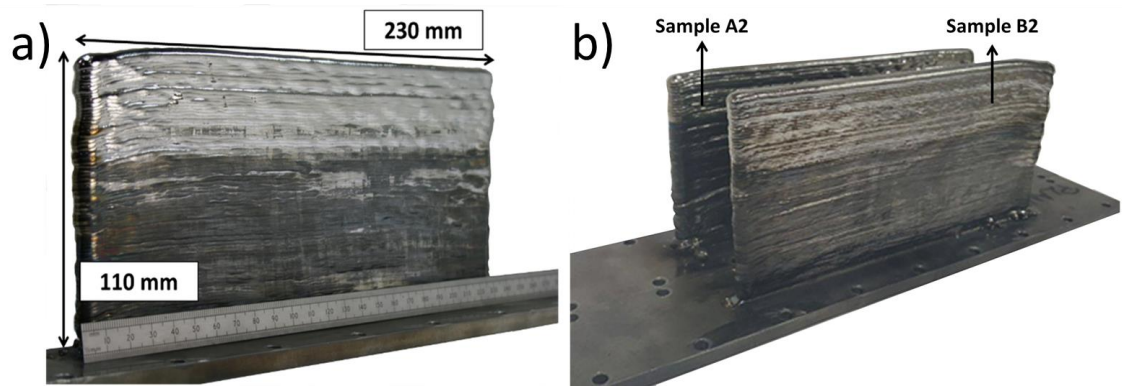

**Fig. 3**: (a) Linear structure made of unalloyed tantalum with the main dimensions; (b) Picture of sample A2 and B2.

**Fig. 4c** and **Fig. 4d** show the results of the characterisation of this outer thin layer, which is around 1.7-µm-thick on average in the lower part of the deposit (**Fig. 4c**). The upper region showed no outer film at all (**Fig. 4d**). This is likely to be an oxide accumulation on the surface but the reason why this occurs is not clear. It is consistent with what seen elsewhere in other WAAM deposits, f.i. in Ti64 structures also produced in an inert environment [24]. Thankfully this feature is not detrimental, because WAAM being a near-net-shape AM process, it typically requires a finish-machining pass, resulting in the elimination of this thin outer film.

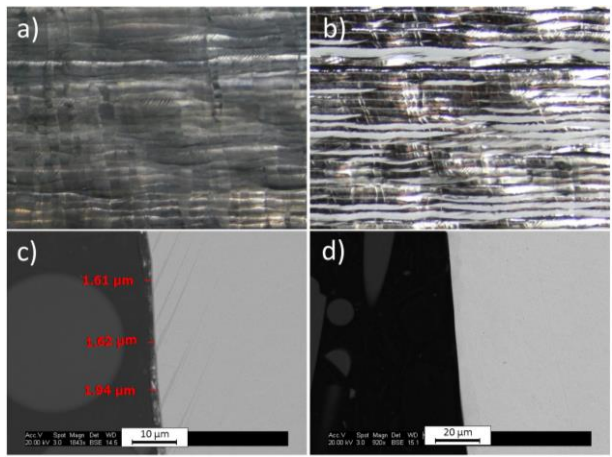

**Fig. 4**: Detail of the outer surface of the bottom (a) and the top (b) of the tantalum structure; Scanning electron microscope picture of the cross-section of the bottom (c) and the top (c) of the structure.

*3.3    Porosity*

**Fig. 5** shows the cross-section of A2 and B2 after polishing. The surface of A2 presented some scattered porosity throughout its height (**Fig. 5a**). In contrast, there was no porosity at all within B2 (**Fig. 5b**). The largest pores had an average diameter of 200 μm; instead, the scattered porosity had an average diameter of 80 μm. As can be seen in **Table 4**, the wire used for the deposition of A2 had a higher concentration of oxygen than the wire used for B2. Therefore, it stands to reason to explain the higher levels of porosity seen in A2 with the higher content of $O_2$ in Wire A. The occurrence and formation of the porosity are discussed later.

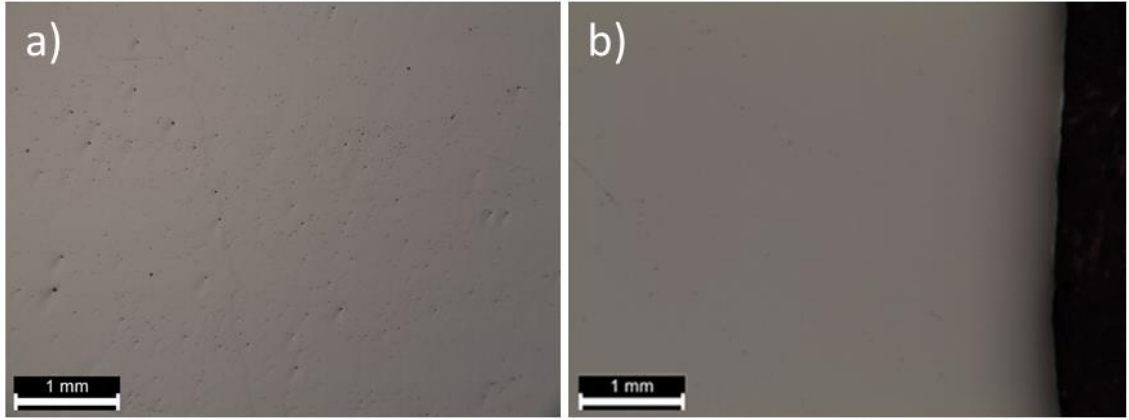

**Fig. 5**: Optical microscope picture of a section of the A2 (a) and B2 (b) after polishing.

*3.4  Microstructure and banding*

**Fig. 6** shows the typical microstructure of the as-deposited tantalum structures produced using WAAM. **Fig. 6a** shows the microstructure of A1, including the interface between substrate and wall. The grain structure of A2 and B2 is shown in **Fig. 6b** and **Fig. 6c**. It can be seen that columnar grains developed epitaxially during the solidification and grew in the direction parallel to the building one, for each of the structures deposited. This is typical for most materials deposited using an AM process. Thijs et al. reported the evolution of columnar grain for tantalum processed by LPBF [20] and the same grain development can be found for Ti-6Al-4V structures processed using WAAM [25]. In particular, the columnar grains increased in size with the height of the wall. Given that an identical grain structure can be seen in both A2 and B2, the different chemistry of the wires plays no role in the crystals' solidification behaviour.

All structures present a macroscopic banding between each deposited layer. Similar cases have been seen for Ti-6Al-4V [9]. In the case of titanium, the banding is due to a localised change of the α-grains due to the re-heating when depositing the successive layer [9,26]. In the case of unalloyed tantalum, secondary phases or sub-grain structures did not develop when exposed to the cyclical re-heating typical of WAAM.

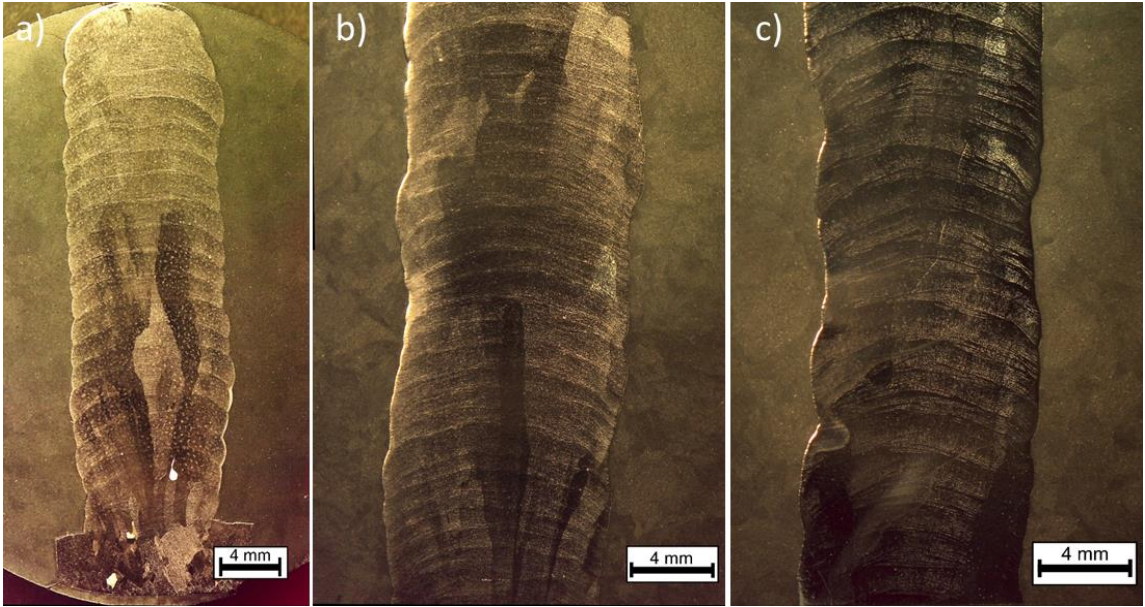

**Fig. 6**: Microstructure of the sample A1 (a), A2 (b) and B2 (c).

**Fig. 7** shows the nature of the banding in A2 between two consecutive layers, which was also observed in B2. Scattered void-like structures were distributed along a line (band) across the interface between two successive layers (**Fig. 7a**). These voids were remarkably smaller than the porosity in A2 (**Fig. 5a**). **Fig. 7b** shows how these features were lined up on successive and almost equidistant line. Additionally, few larger pores were also found within the banding regions. Noteworthy is also the appearance of the banding regions only after etching. As **Fig. 5** shows, none of these regions was observed after polishing only. A reasonable hypothesis is that these void-like features forming the bands are caused by the detachment, caused by the etching, of fine oxide particles, which possibly formed during the solidification of the liquid pool. The average banding distance resulted to be similar to the layer height for all the structures. Thus, this phenomenon seemed to be directly connected to the solidification of each layer.

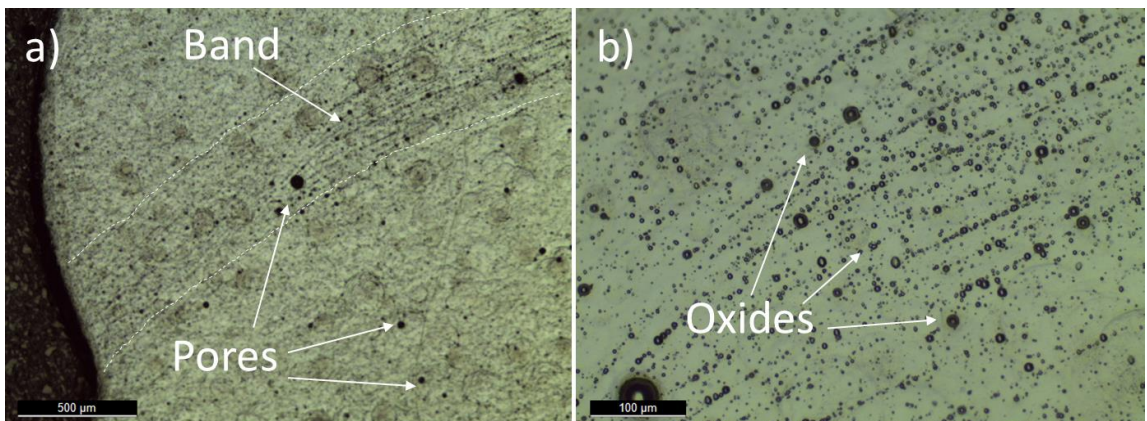

**Fig. 7**: Banding, pores (a) and void-like structures within the band (b) in the sample A2.

Oxygen may be present in two chemical states within a tantalum component: in solid-solution or as oxide precipitate [27]. Their occurrence depends on process temperature, cooling rate and oxygen concentration [27]. The oxygen in solid-solution mainly impacts the average hardness of the material (as it will be reported later in this study). The oxide precipitates are volatile above 1600°C [28]; some studies reported the formation of these particles even at very low concentrations of oxygen [29]. Furthermore, they are strongly attacked by the etchant, which exposes features similar to voids [27]. In fact, Stecura et al. [29] also found voids with the same morphology and features of the voids in **Fig. 7b**, not only at the grain boundaries but also within the grains.

Therefore, the bands could be caused by the presence of a regular dispersion of tantalum oxides particles, in particular $Ta_2O_5$, which is the most stable compound [30]; they were visible due to the strong attack by the etchant because of the high reactivity of these regions. Furthermore, the dispersed porosity, seen in **Fig. 5**, could have been caused by the vaporization of oxide particles when temperatures as high as 3000°C were reached at the bottom of the weld pool while depositing the successive layer. A complete understanding of this phenomenon will be the object of future studies.

*3.5    Hardness*

The average hardness was evaluated for the Substrate, A2 and B2; instead, A1 and B1 have been used to evaluate the hardness at the substrate-wall interface. For all the measurements, hardness values have been measured along the centreline of the structure. In particular, the values for A2 and B2 have been measured along their height between 60.0 mm and 70.0 mm from the substrate. **Table 5** reports the average hardness values for each condition and correlated with the oxygen content.

**Table 5**: The average hardness value of the Substrate, A2 and B2 with the associated standard deviation and the oxygen content in ppm.

|  | Hardness [HV] | Standard Deviation | Oxygen content [ppm] |
|---|---|---|---|
| **Substrate** | 96.0 | 1.3 | 60 |
| **A2** | 114 | 0.8 | 226 |
| **B2** | 99.0 | 1.0 | 164 |

The hardness shows an obvious trend when correlated to the oxygen content of the three samples. The higher the content of oxygen, the higher the resulting average hardness value. This agrees with what found in the literature [19]; in particular, nitrogen and oxygen are strong hardeners of unalloyed polycrystalline tantalum. In the work of Stecura [29], the hardness of tantalum increased almost linearly when increasing oxygen content in solid solution.

Furthermore, it has been reported that the hardness of unalloyed tantalum is mostly governed by the content of oxygen in solid solution and, contrarily, the oxide precipitates have no considerable effect due to their incoherence with the crystals [27]. Spitzig et al. [31] reported about the solid solution hardening due to oxygen as interstitial for vanadium, niobium and tantalum. The effect has been associated with the misfit between the oxygen and the metal atoms. The different average hardness values for the deposited walls (A2 and B2) can be attributed mainly to the oxygen content in solid solution because the grain structure was almost identical, and no large variation of hardness has been measured at the band regions.

     **Fig. 8** shows the variation of the hardness from the bottom of the substrate to the top of A1 and B1. The origin of the X-axis has been chosen as the top surface of the substrate. From **Fig. 8**, it is possible to underline three different regions of interest: Substrate, Zone 1 and Zone 2. The average hardness values for each region and for both A1 and B1 are reported in **Table 6**.

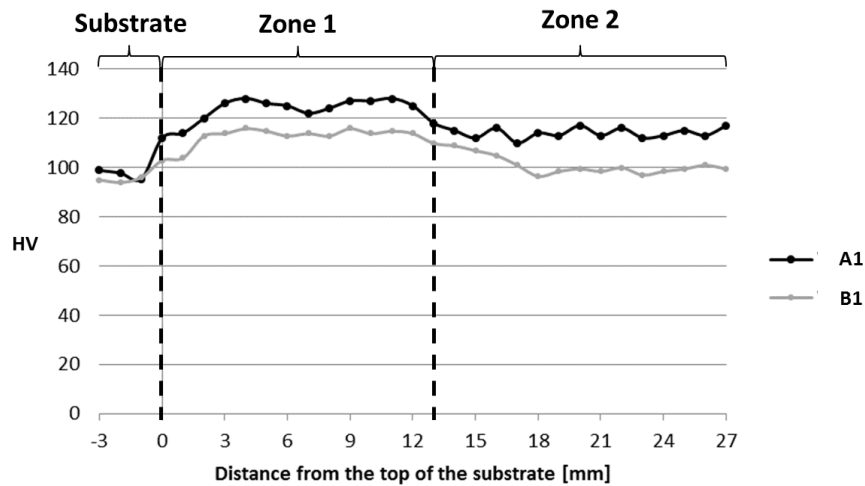

**Fig. 8**: Hardness values of the tantalum A1 and B1 with respect to the distance from the top of the substrate.

**Table 6**: Average hardness within Substrate, Zone 1 and Zone 2 for both sample A1 and B1 with the associated standard deviation.

|  | Substrate | A1 Zone 1 | A1 Zone 2 | B1 Zone 1 | B1 Zone 2 |
|---|---|---|---|---|---|
| **Average** | 96 | 123 | 114 | 112 | 100 |
| **Stand. Dev.** | 1.15 | 5 | 2 | 4 | 3.5 |

     The average hardness of Zone1 was measured to be always higher than the hardness of Zone2 for both A1 and B1. This difference can be possibly addressed to the considerable effect of thermal straining, which in turns induced strain hardening on the lower part of the deposit. Unalloyed polycrystalline tantalum is

known to be subject to great work hardening when a consistent amount of stress is applied [32]. When pure materials are subjected to cyclical deformation, they exhibit an initial hardening which stabilises for any further deformations [33]. In general, for a soft pure metal, the dislocation density is low but, after cyclic plastic straining, the dislocation density increases rapidly causing a strengthening effect called cycling hardening. This is confirmed by comparing the values of the average hardness in **Table 5** and the hardness of Zone2 in **Table 6** for both walls.

*3.6   Tensile properties*

In **Table 7** the main tensile properties for Substrate, A2 and B2 in both vertical (V) and horizontal (H) directions are summarised. **Fig. 9** shows the representative stress-strain curves for some of these conditions. It is evident that for both deposited structures and for both testing directions, a higher yield strength compared to the Substrate has been measured for the WAAM deposits. Contrarily, a clear reduction in total elongation can be seen for both A2 and B2 with respect to the Substrate. Additionally, an evident anisotropy has been observed for both walls.

**Table 7**: Tensile properties of the Substrate, A2 and B2 in both vertical and horizontal direction.

| Coupons Denomination | Yield Strength [MPa] | St. Dev. | Ultimate Tensile Strength [MPa] | St. Dev. | Elongation [%] | St. Dev. |
|---|---|---|---|---|---|---|
| Substrate | 182 | 3.7 | 265 | 7.8 | 62 | 2.5 |
| A2 V | 216 | 10.0 | 241 | 10.0 | 27 | 1.0 |
| A2 H | 234 | 7.0 | 261 | 3.0 | 36 | 6.0 |
| B2 V | 194 | 1.9 | 218 | 8.2 | 35 | 3.0 |
| B2 H | 208 | 14 | 226 | 9.8 | 32 | 7.0 |

The major contribution to the high yield strength of the deposited structures was given by the higher content of oxygen. As already reported in the literature, the yield stress increases linearly with increasing amounts of nitrogen or oxygen [19]. This can also be seen when comparing the yield strength of A2 over B2 in both testing directions (**Fig. 9a**).

The impurity level in tantalum can have a much higher influence on the mechanical strength compared to the grain size [18,34]. Interstitials hinder the propagation of dislocations, increasing the absolute stress for plastic deformation [34]. The large reduction in elongation can be explained mainly by the differences in the size and shape of grains, between the substrate and the deposited structures: the former was characterised by fine equiaxed microstructure, unlike the latter which showed large columnar grains. For a relatively small average grain size under uniaxial load, grain rotation and grain sliding become likely [35,36]. The high mobility of the fine equiaxed grains has possibly confined or stopped the formation and growth of the voids within the gauge volume. The strain accumulation for both substrate and WAAM deposits is discussed later.

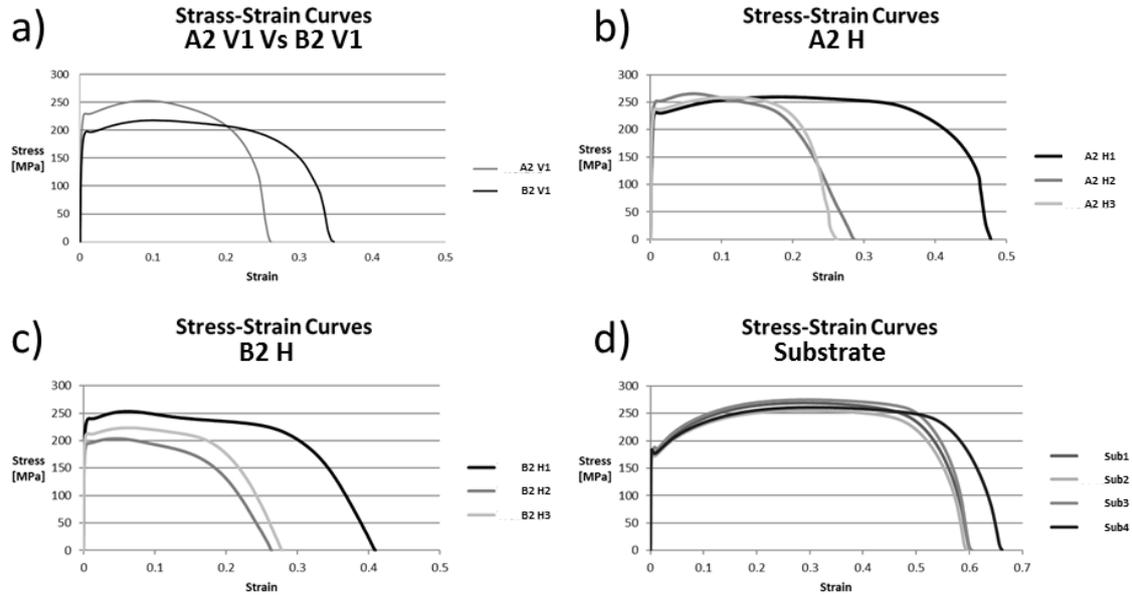

**Fig. 9**: Representative stress-strain curves of Substrate (a), A2 and B2 in the vertical direction (b), A2 (c) and B2 (d) in the horizontal direction.

The yield strength and the ultimate tensile strength were higher for the horizontal samples compared to the vertical ones, for both structures. A similar scenario has been reported for Ti-6Al-4V deposited using Wire + Arc and Wire + Laser Additive Manufacture[9,37]. Baufeld et al. [38] also reported a consistent anisotropy when testing vertical and horizontal samples extracted from an AM component. Thijs et al. [20] reported a marked anisotropy with regards the mechanical properties for tantalum structure deposited using LPBF. The anisotropy has been directly connected to crystallographic texture, grain shape and size developed during the deposition of the directional columnar growth. In particular, considering the long columnar grains with high aspect ratio, the grain size within the gauge volume for the horizontal direction was smaller on average than in the vertical direction. The higher amount of grain boundaries influenced the tensile properties considerably [38].

From **Fig. 9b** and **Fig. 9c**, it is possible to see that the horizontal coupons extracted from the top of the structures (A2 H1 and B2 H1) presented a larger elongation with respect to the coupons extracted from the root of the deposits (A2 H3 and B2 H3). The effect related to the location of the coupon within the structure has been attributed to the thermal straining, as already discussed above for the hardness.

*3.7 Strain Localisation*

**Fig. 10** shows the strain maps under tensile load at different global displacement. A consistent scale ranging from 0 mStrain and 500 mStrain was used in all cases, to facilitate the comparison. The distance covered after a specific time by the cross-head of the tensile machine was used as global displacement and is reported in millimetres. The strain maps shown represent the evolution of the strain from the beginning of the test until about uniform elongation. After this point, the speckle pattern was damaged by the heavy necking and strain values could not be acquired.

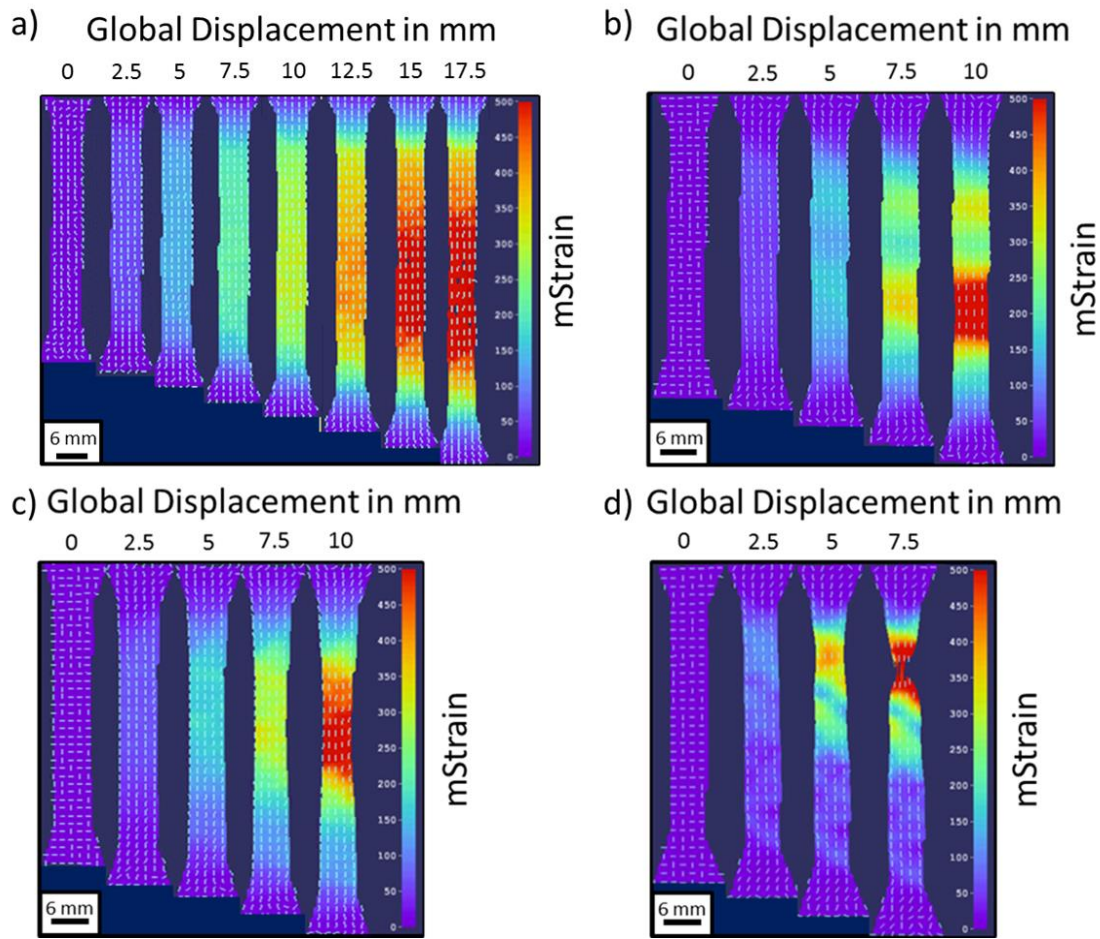

**Fig. 10**: Strain maps produced with digital image correlation for (a) substrate, (b) vertical coupon from A2 (A2 V2), (c) horizontal coupon close to the top of the B2 (B2 H1) and (d) horizontal coupon close to the bottom of B2 (B2 H3).

The tensile coupons extracted from the walls presented a faster strain localisation compared to those extracted from the substrate. This is due to the difference in the microstructure. As discussed previously, for finer microstructures the mobility of the grains under uniaxial load is high. The re-orientation and sliding of the grains within the substrate led to a much more uniform distribution of the strain through the gauge length. For the vertical specimen A2 V2, the formation of voids happened predominantly within the large grains leading to a high concentration of strain and a much smaller uniform-elongation volume. Furthermore, the fracture and the formation of micro-voids have possibly been enhanced by the presence of the oxide particles forming the bands. As already reported, the coupon in **Fig. 10c** and **Fig. 10d** had similar values of yield strength and ultimate tensile strength but a large difference in elongation. The sudden localisation of the strain for the sample B2 H3 (**Fig. 10d**) corroborates the explanation reported before for the discrepancies in elongation. It appears that the abrupt strain localisation has been enhanced by a percentage of strain already present before the test for the cyclic straining effect.

## 4    Conclusions

In this research, the Wire + Arc Additive Manufacturing process has been studied and proven for tantalum. The chemistry of two different unalloyed tantalum wires has been correlated to the microstructure, porosity and mechanical properties of the as-deposited structure. The main findings of this study can be summarised as follows:

- An anisotropic microstructure has been found. Large columnar grains grew epitaxially from the substrate through the entire height of the deposit. The presence of macroscopic bands, given by small craters caused by the removal of dispersed fine oxide particles, was also observed at the edge of each deposited layer;

- The content of oxygen within the two different wires was one of the main drivers influencing porosity and average hardness. In particular, a larger content of oxygen led to the development of micron-size porosity and higher hardness values;

- The tensile properties of the deposited walls were comparable to those of the substrate material, unlike the total elongation. The high yield strength of the walls has been explained by the material's chemistry and in particular by the content of oxygen. The loss in elongation is related to the shape and size of grains;

- Thermal straining and repetitive thermal cycles due to the deposition of the successive layers influenced the deposited structures markedly. A cyclic hardening effect has been seen at the substrate-wall interface by measuring the hardness profile. A cyclic plastic strain has also influenced the mechanical properties along the height of the structure for the horizontal samples.

In conclusion, WAAM has proven capable of depositing high-integrity tantalum structures, with excellent integrity, low porosity and satisfactory mechanical properties, provided a wire of good quality is used as feedstock. The reduced cost associated with WAAM processing might open-up new applications for this very interesting element.

**Acknowledgement**

The authors wish to acknowledge financial support from AWE and the valuable scientific contribution of Geoff Shrimpton (AWE) and Tim Rogers (AWE).